\documentclass[conference,romanappendices]{IEEEtran}
\IEEEoverridecommandlockouts

\usepackage{algorithm,algorithmic,amsmath,amssymb,amsthm,bbm,cite,color,graphicx,microtype,url}
\usepackage[USenglish]{babel}
\usepackage{hyperref}
\usepackage[utf8]{inputenc}
\usepackage[T1]{fontenc}
\usepackage{subcaption}
\usepackage[figurename=Fig., labelsep=period, font=small]{caption}

\usepackage{tikz,pgfplots}
\usetikzlibrary{arrows}
\pgfplotsset{compat=1.17}

\usepackage{titlesec}
\let\subparagraph\relax
\titlespacing{\section}{0pt}{6pt plus 2pt minus 1pt}{4pt plus 1pt minus 1pt} % space above and below section titles
\titlespacing{\subsection}{0pt}{4pt plus 2pt minus 1pt}{2pt plus 1pt minus 1pt} % space above and below subsection titles
\renewcommand{\d}{\mathbf{d}}

\newcommand{\f}{\mathbf{f}}

\newcommand{\n}{\mathbf{n}}

\newcommand{\p}{\mathbf{p}}
\newcommand{\q}{\mathbf{q}}

\newcommand{\s}{\mathbf{s}}

\newcommand{\x}{\mathbf{x}}
\newcommand{\y}{\mathbf{y}}
\newcommand{\z}{\mathbf{z}}

\newcommand{\0}{\mathbf{0}}

\newcommand{\A}{\mathbf{A}}

\newcommand{\C}{\mathbf{C}}

\newcommand{\F}{\mathbf{F}}
\newcommand{\G}{\mathbf{G}}
\renewcommand{\H}{\mathbf{H}}
\newcommand{\I}{\mathbf{I}}

\renewcommand{\P}{\mathbf{P}}

\newcommand{\V}{\mathbf{V}}
\newcommand{\W}{\mathbf{W}}
\newcommand{\X}{\mathbf{X}}

\newcommand{\Z}{\mathbf{Z}}

\newcommand{\mub}{\boldsymbol{\mu}}

\newcommand{\setC}{\mathcal{C}}

\newcommand{\setN}{\mathcal{N}}

\newcommand{\setQ}{\mathcal{Q}}

\newcommand{\setS}{\mathcal{S}}

\newcommand{\Compl}{\mbox{$\mathbb{C}$}}

\newcommand{\argmin}{\operatornamewithlimits{argmin}}

\newcommand{\Diag}{\mathrm{Diag}}
\newcommand{\diff}{\mathrm{d}}

\newcommand{\herm}{\mathrm{H}}
\renewcommand{\Im}{\mathrm{Im}}
\newcommand{\logdet}{\mathrm{logdet}}

\renewcommand{\Re}{\mathrm{Re}}
\newcommand{\sgn}{\mathrm{sgn}}

\newcommand{\tran}{\mathrm{T}}

\definecolor{oulu_blue}{HTML}{23408F}
\definecolor{oulu_green}{HTML}{39B54A}

\definecolor{red}{rgb}{1,0,0}
\definecolor{red_magenta}{rgb}{1,0,0.5}
\definecolor{magenta}{rgb}{1,0,1}
\definecolor{blue_magenta}{rgb}{0.5,0,1}
\definecolor{blue}{rgb}{0,0,1}
\definecolor{blue_cyan}{rgb}{0,0.5,1}
\definecolor{cyan}{rgb}{0,1,1}
\definecolor{green_cyan}{rgb}{0,1,0.5}
\definecolor{green}{rgb}{0,1,0}
\definecolor{green_yellow}{rgb}{0.5,1,0}
\definecolor{yellow}{rgb}{1,1,0}
\definecolor{red_yellow}{rgb}{1,0.5,0}

\title{Massive MIMO with 1-Bit DACs: Data Detection for Quantized Linear Precoding with Dithering}
\author{
\IEEEauthorblockN{Amin Radbord, Italo Atzeni, and Antti Tölli}
\IEEEauthorblockA{Centre for Wireless Communications, University of Oulu, Finland \\
E-mail: \{amin.radbord, italo.atzeni, antti.tolli\}@oulu.fi}
\thanks{This work was supported by the Research Council of Finland (336449 Profi6, 348396 HIGH-6G, 357504 EETCAMD, and 369116 6G~Flagship) and by the European Commission (101095759 Hexa-X-II).}}

\begin{document}

\maketitle

\begin{abstract}
To leverage high-frequency bands in 6G wireless systems and beyond, employing massive multiple-input multiple-output (MIMO) arrays at the transmitter and/or receiver side is crucial. To mitigate the power consumption and hardware complexity across massive frequency bands and antenna arrays, a sacrifice in the resolution of the data converters will be inevitable. In this paper, we consider a point-to-point massive MIMO system with 1-bit digital-to-analog converters at the transmitter, where the linearly precoded signal is supplemented with dithering before the 1-bit quantization. For this system, we propose a new maximum-likelihood (ML) data detection method at the receiver by deriving the mean and covariance matrix of the received signal, where symbol-dependent linear minimum mean squared error estimation is utilized to efficiently linearize the transmitted signal. Numerical results show that the proposed ML method can provide gains of more than two orders of magnitude in terms of symbol error rate over conventional data detection based on soft estimation.
\end{abstract}

\begin{IEEEkeywords}
1-bit DACs, massive MIMO, maximum-likelihood data detection.
\end{IEEEkeywords}

%=========================================================================
\section{Introduction} \label{sec:Intro}
%=========================================================================

Deploying massive multiple-input multiple-output (MIMO) systems is crucial to leverage the wide bandwidths available at high frequencies for 6G and future wireless networks \cite{Raj20}. To realize massive MIMO arrays, fully digital architectures are preferred since they provide highly flexible beamforming and large-scale spatial multiplexing without the need for the complex beam-management schemes of their hybrid analog-digital counterparts. However, the power consumption of the analog-to-digital/digital-to-analog converters (ADCs/DACs) scales linearly with the bandwidth and exponentially with the number of resolution bits \cite{Atz21b}. This issue has encouraged research into the implementation of low-resolution ADCs/DACs. Simple 1-bit ADCs/DACs are especially appealing thanks to their minimal power consumption~\cite{Li17,Mol17,Atz23}, so they can be easily deployed in very large numbers to create massive arrays. In particular, 1-bit DACs relax the requirements for high-cost radio-frequency components at the transmitter and enable very high energy efficiency since the power amplifiers can operate without back-off.

There is a vast literature on quantized precoding design with 1-bit DACs \cite{Jac17b,Sax17,AnLi17,Lian23,Usm17}. Symbol-level precoding (SLP) based on, e.g., constructive interference and constructive region \cite{Sax17,AnLi17} or the minimum mean squared error (MMSE) criterion \cite{Lian23}, achieves very good performance at the cost of considerable complexity. On the other hand, quantized linear precoding offers a good balance between performance and complexity. In this context, the linear precoding matrix is optimized for doubly 1-bit quantized systems in \cite{Usm17}. However, this work considers additional analog gains after the 1-bit quantization and the analysis involves multiple approximations. Furthermore, \cite{Sax19,Sax20} apply Gaussian dithering to the linearly precoded signal before the 1-bit quantization. In this respect, an optimized dithering power can reduce the strong correlation among the quantization distortion components across the transmit antennas and thus enable the recovery of amplitude information, e.g., from quadrature amplitude modulation (QAM) symbols. The benefits of dithering for quantized signals are widely recognized \cite{Rap19}, especially in the context of 1-bit ADCs \cite{Atz22}.

In this paper, we focus on the data detection in a point-to-point massive MIMO system with 1-bit DACs at the transmitter, where the linearly precoded signal is supplemented with Gaussian dithering before the 1-bit quantization. Instead of considering data detection via linear combining based on the MMSE criterion as in \cite{Sax19,Sax20}, we propose a new maximum-likelihood (ML) data detection method at the receiver. In this regard, we present an analytical framework by deriving the statistics of the received signal (i.e., the mean and covariance matrix), where symbol-dependent linear MMSE (LMMSE) estimation is utilized to efficiently linearize the transmitted signal. We also extend \cite{Sax19,Sax20} to multiple receive antennas and data streams. We then investigate the impact of the dithering power, the number of receive antennas and data streams, and the signal-to-noise ratio (SNR) on the system's performance. Numerical results show that the proposed ML method can provide gains of more than two orders of magnitude in terms of symbol error rate (SER) compared with conventional data detection based on soft estimation.

%=========================================================================
\section{System Model} \label{sec:Sys}
%=========================================================================

Consider a point-to-point massive MIMO system where a transmitter equipped with $N$ antennas and 1-bit DACs transmits $K$ data streams to a receiver with $M$ antennas and full-resolution ADCs, with $K \leq \min(N, M)$. This setup may represent, for instance, a wireless backhaul scenario, as considered in Section~\ref{sec:NR}. To model the 1-bit DACs at the transmitter, we introduce the 1-bit quantization function $Q(\cdot) : \Compl^{A} \rightarrow \setQ$, with $\setQ \triangleq \sqrt{\frac{\eta}{2}} \{ \pm 1 \pm j \}^{A}$ and

$ $ \vspace{-8mm}

\begin{align} \label{eq:Q}
Q(\x) \triangleq \sqrt{\frac{\eta}{2}} \big( \sgn( \Re[\x]) + j \, \sgn (\Im[\x]) \big),
\end{align}
with scaling factor $\eta > 0$. Let \( \H \in \mathbb{C}^{M \times N} \) represent the channel matrix between the transmitter and the receiver. In this work, we assume perfect channel state information (CSI) is available at both the transmitter and receiver, while the analysis under imperfect CSI is left for future work.

The transmitter aims at delivering the data symbol vector $\s \in \mathbb{C}^{K}$ to the receiver. To do so, the transmitter employs a quantized linear precoding strategy with properly designed dithering, as in \cite{Sax19,Sax20}. First, the precoding matrix $\W$ is computed based on $\H$ and independently of $\s$. Then, the resulting signal $\x \triangleq \W \s \in \Compl^{N}$ is supplemented with the Gaussian dithering vector $\d \sim \mathcal{CN}(\0_{N}, \sigma^2 \I_N)$, which helps mitigate the effects of coarse quantization before the signaling is processed by the 1-bit DACs. Specifically, the Gaussian dithering can reduce the strong correlation among the quantization distortion components across the transmit antennas, particularly when the number of data streams is small. Given the dithered precoded signal $\x_{\textrm{d}}\triangleq \x + \d \in \mathbb{C}^{N}$, the transmitted signal (after the 1-bit DACs) is given by
\begin{align}\label{eq:xq}
    \x_{\textrm{q}} \triangleq Q(\x_{\textrm{d}}) \in \Compl^{N},
\end{align}
where the scaling factor $\eta$ of the 1-bit quantization function in \eqref{eq:Q} is fixed as $\eta = \frac{1}{N}$ to satisfy the power constraint $\|\x_{\textrm{q}}\|^2 = 1$.

Subsequently, the analog signal $\x_{\textrm{q}}$ is transmitted over the channel with transmit power $\rho$. The signal arriving at the receiver is given by
\begin{align}\label{eq:y}
    \y & \triangleq \sqrt{\rho}\H \x_{\textrm{q}} + \z \in \mathbb{C}^{M},
\end{align}
where $\z \sim \mathcal{CN}(\0_{M}, \I_M)$ is a vector of additive white Gaussian noise (AWGN). Since the AWGN is assumed to have unit variance, $\rho$ can be interpreted as the transmit SNR. Finally, the receiver may obtain a soft estimate of $\s$ via linear combining of the received signal $\y$ as
\begin{equation}\label{eq:shat}
    \hat{\s} \triangleq \V^\herm \y \in \mathbb{C}^{K},
\end{equation}
where $\V \in \mathbb{C}^{M \times K}$ is the combining matrix. In this paper, we propose an alternative data detection strategy that does not rely on the soft-estimated symbols but instead directly detects the data symbol vector $\s$ from the received signal $\y$ in \eqref{eq:y}.

%=========================================================================
\section{Linearization of the Transmitted Signal and Proposed ML Data Detection} \label{sec:LQP&DD}
%=========================================================================

The coarse quantization from the 1-bit DACs at the transmitter introduces a non-linear distortion due to the discrete nature of the quantization process. To obtain a tractable framework, we first linearize the transmitted signal resulting from the 1-bit quantization of the dithered precoded signal. In this regard, we introduce two linearization methods: 1) linearization via Bussgang decomposition, which is used to design a linear combining matrix for the data detection based on soft estimation that is considered for comparison; and 2) linearization via symbol-dependent LMMSE estimation, which is employed to establish the proposed data detection method presented in Section~\ref{sec:LQP&DD_D}.

%=========================================================================
\subsection{Linearization via Bussgang Decomposition} \label{sec:LQP&DD_A}
%=========================================================================

In this section, we express the transmitted signal as a linear function using the Bussgang decomposition \cite{Dem21}, which allows one to write the output of a nonlinear system as the sum of a scaled version of the input and an uncorrelated distortion term. To exploit the Bussgang decomposition, we assume $\x_{\textrm{d}}$ to be Gaussian, i.e., $\x_{\textrm{d}} \sim \setC \setN (\0_{N}, \C_{\x_{\textrm{d}}})$. Hence, we consider Gaussian data symbols, i.e., $ \s \sim \mathcal{CN}(\0_{K}, \mathbf{I}_K)$. Let us define
\begin{align}
    \C_{\x_{\textrm{d}}} \triangleq \mathbb{E}_{\s,\d}[\x_{\textrm{d}}\x_{\textrm{d}}^\herm] = \W\W^\herm + \sigma^2\I_N \in \Compl^{N \times N}.
\end{align}
We linearize $\x_{\textrm{q}}$ in \eqref{eq:xq} with respect to $\x_{\textrm{d}}$ (and, thus, with respect to $\s$ and $\d$) using the Bussgang decomposition as
\begin{align}\label{eq:xq (1)}
    \x_{\textrm{q}} = \F\x_{\textrm{d}} + \q_{\textrm{d}},
\end{align}
where $\q_{\textrm{d}}$ is a zero-mean, non-Gaussian distortion vector that is uncorrelated with $\x_{\textrm{d}}$ and $\F$ is a diagonal matrix that can be computed in closed form as \cite{Li17}
\begin{align}\label{eq:B}
    \F \triangleq \sqrt{\frac{2}{\pi}\eta}\Diag(\C_{\x_{\textrm{d}}})^{-\frac{1}{2}} \in \mathbb{C}^{N\times N}.
\end{align}

\begin{figure*}[t!]
\setcounter{equation}{10}
\begin{align}\label{eq:Cxdxq|x (1)}
[\C_{\x_{\textrm{d}}\x_{\textrm{q}}|\x}]_{n,m} &= [\C_{\x_{\textrm{d}}\x_{\textrm{q}}|\x}^{( \Re, \Re)}]_{n,m} + [\C_{\x_{\textrm{d}}\x_{\textrm{q}}|\x}^{( \Im, \Im)}]_{n,m} - j\,[\C_{\x_{\textrm{d}}\x_{\textrm{q}}|\x}^{( \Re, \Im)}]_{n,m} + j\,[\C_{\x_{\textrm{d}}\x_{\textrm{q}}|\x}^{( \Im, \Re)}]_{n,m} \\
& = \begin{cases}
    \sqrt{\frac{\eta}{2}} \Big( \sqrt{\frac{\sigma^2}{\pi}} \Big(\exp \big( \! - \! \big(\frac{\Re[x_m]}{\sigma}\big)^2 \big)  \! + \! \exp \big( \! - \! \big(\frac{\Im[x_m]}{\sigma} \big)^2 \big)\Big) \! + \! \Re \big[x_n\Phi_{\textrm{c}} \big(\frac{x_m^*}{\sigma} \big) \big]  \! + \! j \, \Im \big[x_n\Phi_{\textrm{c}} \big(\frac{x_m^*}{\sigma} \big) \big] \Big) & \ \textrm{if}~m=n, \\
  \sqrt{\frac{\eta}{2}} \big(\Re \big[x_n\Phi_{\textrm{c}} \big(\frac{x_m^*}{\sigma} \big) \big] +j \, \Im \big[x_n\Phi_{\textrm{c}} \big(\frac{x_m^*}{\sigma} \big) \big] \big) & \ \textrm{otherwise},
    \end{cases}\label{eq:Cxdxq|x (2)} \\
    \label{eq:Exd_n Exq_m (App)}
    [\C_{\x_{\textrm{d}}\x_{\textrm{q}}|\x}^{( \mathrm{A}, \mathrm{B})}]_{n,m} &\triangleq \sqrt{\frac{\eta}{2}}\mathbb{E} \big[\mathrm{A}[x_{\textrm{d},n}]\,\sgn \big(\mathrm{B}[x_{\textrm{d},m}] \big)|\x\big] =
    \begin{cases}
    \sqrt{\frac{\eta}{2}}\Big(\sqrt{\frac{\sigma^2}{\pi}} \exp \big(\! - \! \big(\frac{\mathrm{A}[x_m]}{\sigma} \big)^2 \big)\delta_{\mathrm{A},\mathrm{B}} \! + \! \mathrm{A}[x_n]\Phi \big(\frac{\mathrm{B}[x_m]}{\sigma} \big)\Big) & \ \textrm{if}~m=n, \\
    \sqrt{\frac{\eta}{2}}\mathrm{A}[x_n]\Phi \big(\frac{\mathrm{B}[x_m]}{\sigma} \big) & \ \textrm{otherwise} 
    \end{cases} 
\end{align}
\hrulefill
\setcounter{equation}{16}
    \begin{align}
   \nonumber & \hspace{-2mm} \C_{\y'|\x} = \\
   & \hspace{-3mm} \begin{bmatrix}
    \sqrt{\rho}\f^{(\Re)}\big( \sqrt{\rho}\f^{(\Re)} \! + \! \mub_{\tilde{\n}}^{(\Re)} \big)^\tran \!\! + \sqrt{\rho}\mub_{\tilde{\n}}^{(\Re)}\big(\f^{(\Re)}\big)^\tran \! +\C_{\tilde{\n}'}^{(\Re,\Re)} \!\!\! & \sqrt{\rho}\f^{(\Re)}\big( \sqrt{\rho}\f^{(\Im)} \! + \! \mub_{\tilde{\n}}^{(\Im)} \big)^\tran \!\! + \! \sqrt{\rho}\mub_{\tilde{\n}}^{(\Re)}(\f^{(\Im)})^\tran \! + \! \C_{\tilde{\n}'}^{(\Re,\Im)} \\
    \sqrt{\rho}\big( \sqrt{\rho}\f^{(\Im)} \! + \! \mub_{\tilde{\n}}^{(\Im)} \big)(\f^{(\Re)})^\tran \!\! + \! \sqrt{\rho}\f^{(\Im)}(\mub_{\tilde{\n}}^{(\Re)})^\tran \! + \! (\C_{\tilde{\n}'}^{(\Re,\Im)})^\tran \!\!\! & \sqrt{\rho}\f^{(\Im)}\big( \sqrt{\rho}\f^{(\Im)} \! + \! \mub_{\tilde{\n}}^{(\Im)} \big)^\tran \!\! + \sqrt{\rho}\mub_{\tilde{\n}}^{(\Im)}\big(\f^{(\Im)}\big)^\tran + \! \C_{\tilde{\n}'}^{(\Im,\Im)}
    \end{bmatrix}\label{eq:Cy'|x}
\end{align}
\hrulefill
\end{figure*}

%=========================================================================
\subsection{Linearization via Symbol-Dependent LMMSE Estimation}\label{sec:LQP&DD_B}
%=========================================================================

In this section, we assume that the data symbol vector $\s$ (and, consequently, $\mathbf{x}$) is fixed, which yields $\mathbf{x}_{\textrm{d}} \sim \mathcal{CN}(\mathbf{x}, \sigma^2 \mathbf{I}_N)$. We linearize $\mathbf{x}_{\textrm{q}}$ in \eqref{eq:xq} for a given $\mathbf{x}$ as
\setcounter{equation}{7}
\begin{align}\label{eq:xq (2)}
     \x_{\textrm{q}} = \G(\x) \x_{\textrm{d}} + \p_{\textrm{d}},
\end{align}
where $\p_{\textrm{d}} \in \mathbb C^{M}$ is a non-Gaussian distortion vector that is uncorrelated with $\x_{\textrm{d}}$. $\G(\x) \in \mathbb{C}^{N\times N}$ is the LMMSE matrix obtained as
\begin{align}
\label{eq:G (1)} \G(\x) & \triangleq \argmin_{\A(\x)} \mathbb{E}_{\d}\big[ \big\|\x_{\textrm{q}} - \A(\x)\x_{\textrm{d}} \big\|^2 \big| \x \big] \\
\label{eq:G (2)} &= \C_{\x_{\textrm{d}}\x_{\textrm{q}}|\x}^\herm \C_{\x_{\textrm{d}}|\x}^{-1},
\end{align}
where $\C_{\x_{\textrm{d}}|\x} \triangleq \mathbb E_{\d}[\x_{\textrm{d}}\x_{\textrm{d}}^\herm|\x] = \x\x^\herm + \sigma^2\I_N \in \mathbb{C}^{N\times N}$ is the symbol-dependent auto-correlation matrix of $\x_{\textrm{d}}$ and $\C_{\x_{\textrm{d}}\x_{\textrm{q}}|\x} \triangleq \mathbb{E}_{\d}[\x_{\textrm{d}}\x_{\textrm{q}}^\herm|\x] \in \mathbb{C}^{N\times N}$ is the symbol-dependent cross-correlation matrix between $\x_{\textrm{d}}$ and $\x_{\textrm{q}}$, which is given in \eqref{eq:Cxdxq|x (1)}--\eqref{eq:Cxdxq|x (2)} at the top of the next page, with specific term in \eqref{eq:Exd_n Exq_m (App)}. Throughout the paper, $\Phi(x) \triangleq \frac{2}{\sqrt{\pi}} \int_{0}^{x} e^{-t^2} \diff t$ represents the error function for a real input $x$; for a complex input $z$, we have $\Phi_{\textrm{c}}(z) \triangleq \Phi \big(\Re[z] \big) +j \, \Phi \big(\Im[z] \big)$. In addition, $x_{n}$ and $x_{\textrm{d},n}$ denote the $n$th elements of $\x$ and $\x_{\textrm{d}}$, respectively. Lastly, $\mathrm{A}[\cdot]$ and $\mathrm{B}[\cdot]$ can be either $\Re[\cdot]$ or $\Im[\cdot]$, whereas $\delta_{\mathrm{A},\mathrm{B}} = 1$ if $\mathrm{A} = \mathrm{B}$ and $\delta_{\mathrm{A},\mathrm{B}} = 0$ otherwise. Moreover, for a complex input matrix $\Z$, $\Lambda_{\mathrm{A},\mathrm{B}}(\Z) \triangleq \delta_{\mathrm{A},\mathrm{B}} \Re[\Z] - (1-\delta_{\mathrm{A},\mathrm{B}})\Im[\Z]$.

%=========================================================================
\subsection{Proposed ML Data Detection}\label{sec:LQP&DD_D}
%=========================================================================

Here, we exploit the symbol-dependent LMMSE estimation in Section~\ref{sec:LQP&DD_B} to formulate a new ML data detection problem that directly detects the data symbol vector $\s$ from the received signal $\y$ in \eqref{eq:y}. We assume $\s \in \setS^{K}$, where $\setS \triangleq \{ s_{1}, \ldots, s_{L} \}$ represents the set of the $L$ possible data symbols. For instance, $\setS$ may correspond to the 16-QAM constellation, as considered in Section~\ref{sec:NR}.

Based on the linearization described in Section~\ref{sec:LQP&DD_B}, we rewrite $\y$ in \eqref{eq:y} as
\setcounter{equation}{13}
\begin{align}\label{eq:y_decomposed}
    \y = \sqrt{\rho}\H \G(\x)\x + \tilde{\n},
\end{align}
where $\tilde{\n} \triangleq \sqrt{\rho}\H \big(\G(\x)\d + \p_{\textrm{d}}\big) + \z \in \mathbb{C}^{M}$ is the effective noise. Let us define $\tilde{\n}' \triangleq \big[\Re[\tilde{\n}],\Im[\tilde{\n}]\big]^\tran \in \mathbb{R}^{2M}$, which is approximately Gaussian when the number of transmit antennas $N$ is large and the elements of $\p_{\textrm{d}}$ are not strongly correlated. This assumption asymptotically holds when the dithering power $\sigma^{2}$ is sufficiently large \cite{Sax19}. However, we observe through numerical simulations that it is already quite accurate for $N\geq 16$ and $\sigma^2 \geq-10$~dBm. Hence, we consider $\tilde{\n}' \sim \mathcal{N} \big(\boldsymbol{\mu}_{\tilde{\n}'}(\x),\mathbf{\Sigma}_{\tilde{\n}'}(\x)\big)$, with mean $\boldsymbol{\mu}_{\tilde{\n}'}(\x) \in \mathbb{R}^{2M}$ and covariance matrix $\mathbf{\Sigma}_{\tilde{\n}'}(\x) \in \mathbb{R}^{2M \times 2M}$ given in \eqref{eq:mu_n'} and \eqref{eq:Sigma_n'}, respectively, in the Appendix.

Let us define $\y' \triangleq \big[\Re[\y], \Im[\y] \big]^\tran \in \mathbb{R}^{2M}$, $\f^{(\mathrm{A})} \triangleq \mathrm{A} \big[\H\G(\x)\x \big] \in \mathbb{R}^{M}$, $\mub_{\tilde{\n}}^{(\mathrm{A})} \triangleq \mathbb{E}\big[\mathrm{A}[\tilde{\n}]\big] \in \mathbb{R}^{M}$, and $\C_{\tilde{\n}'}^{(\mathrm{A,B})} \triangleq \mathbb{E}\big[\mathrm{A}[\tilde{\n}]\mathrm{B}[\tilde{\n}]^\tran\big] \in \mathbb{R}^{M\times M}$ (see the Appendix). From \eqref{eq:y_decomposed}, we have $\y' \sim \mathcal{N}(\boldsymbol{\mu}_{\y'|\x},\mathbf{\Sigma}_{\y'|\x})$, with mean
\begin{align}
    \boldsymbol{\mu}_{\y'|\x} &\triangleq \mathbb{E}[\y'|\x] = \begin{bmatrix}
    \sqrt{\rho} \f^{(\Re)} + \mub_{\tilde{\n}}^{(\Re)}\\
      \sqrt{\rho} \f^{(\Im)} + \mub_{\tilde{\n}}^{(\Im)}
    \end{bmatrix} \in \mathbb{R}^{2M} \label{eq:mu}
\end{align}
and covariance matrix
\begin{align}
    \mathbf{\Sigma}_{\y'|\x} \triangleq \C_{\y'|\x} - \boldsymbol{\mu}_{\y'|\x}\boldsymbol{\mu}_{\y'|\x}^\tran \in \mathbb{R}^{2M\times 2M},
\end{align}
where $\C_{\y'|\x} \triangleq \mathbb{E}[\y'\y'^\tran|\x] \in \mathbb{R}^{2M\times 2M}$ is detailed in \eqref{eq:Cy'|x} at the top of the page. Recalling that $\x = \W\s$, the ML data detection problem is formulated as
\setcounter{equation}{17}
\begin{align}
    \nonumber\hat{\s}_{\textrm{ML}} &= \argmin_{\s \in \setS^{K}} \, ( \y' - \boldsymbol{\mu}_{\y'|\x})^\tran \mathbf{\Sigma}_{\y'|\x}^{-1}( \y' -  \boldsymbol{\mu}_{\y'|\x}) \\
    & \phantom{=} \ + \logdet (\mathbf{\Sigma}_{\y'|\x}).
\end{align}
The computational complexity of the proposed ML data detection method is $\mathcal{O}(NML^K)$, which is feasible for small values of $K$. A low-complexity variant based on near ML approaches (see, e.g., \cite{Cho16}) will be the subject of future work.

In Section~\ref{sec:NR}, we compare the proposed ML method with data detection based on soft estimation of the received signal in \eqref{eq:y}, which utilizes the Bussgang decomposition described in Section~\ref{sec:LQP&DD_A}. For the combining matrix, we consider the Bussgang LMMSE (BLMMSE) receiver, which minimizes the mean squared error (MSE) between $\hat{\s}$ in \eqref{eq:shat} and the data symbol vector $\s$, yielding
\begin{align} \label{eq:VBLMMSE}
    \V_{\textrm{BLMMSE}} = \sqrt{\rho} \C_{\y}^{-1}\H\F\W,
\end{align}
with $\C_{\y} \triangleq \mathbb{E}[\y\y^\herm] \in \mathbb{C}^{M\times M}$ and where $\F$ is defined in \eqref{eq:B}. The proof of \eqref{eq:VBLMMSE} (including the derivation of $\C_{\y}$) follows similar steps as in \cite{Atz23} and is thus omitted due to space limitations. Then, we apply the minimum distance criterion to map each soft-estimated symbol of each stream to one of the elements in $\setS$.

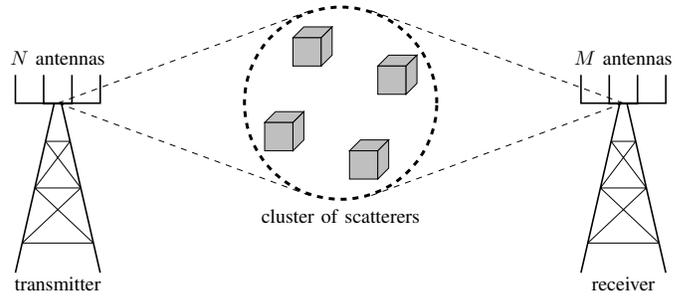
\begin{figure}[t!]
\centering
\resizebox{0.5\textwidth}{!}{\begin{tikzpicture}[pics/power pole/.style={code={
 \draw[thick] \foreach \X in {0.25, 0.75} {(-\X,0.5) |- (\X,0) -- (\X,0.5)}
    (-0.75, -3) -- (-0.0625, 0) coordinate[pos=0.175] (l3)
        coordinate[pos=0.5] (l2) coordinate[pos=0.775] (l1)
    (0.75, -3) -- (0.0625, 0) coordinate[pos=0.175] (r3)
        coordinate[pos=0.5] (r2) coordinate[pos=0.775] (r1);
  \draw (l1) -- (r1) (l2) -- (r2) (l3) -- (r3)
  (l1) -- (r2) (l2) -- (r3) (l3) -- (r2) (l2) -- (r1);
}}]

\path  pic at (0, 3){power pole};
\path  pic at (10, 3){power pole};

    % Draw the dashed circle representing the coverage area
    \draw[dashed,line width=1.5pt] (5,3) circle(1.7); % Dashed circle around the scatterers

    % 3D scatterer cubes inside the dashed circle
% Create the cubes with a 3D perspective
% Scatterer 1
\newcommand{\drawCube}[3]{%
    \draw[black,fill=gray!50,] (#1, #2, 0) -- (#1+0.5, #2, 0) -- (#1+0.5, #2, 0.5) -- (#1, #2, 0.5) -- (#1, #2, 0) -- cycle; % front face
    \draw[black,fill=gray!50] (#1, #2, 0.5) -- (#1+0.5, #2, 0.5) -- (#1+0.5, #2-0.5, 0.5) -- (#1, #2-0.5, 0.5) -- (#1, #2, 0.5) -- cycle; % side face
    \draw[black,fill=gray!50] (#1+0.5, #2, 0.5) -- (#1+0.5, #2, 0) -- (#1+0.5, #2-0.5, 0) -- (#1+0.5, #2-0.5, 0.5) -- (#1+0.5, #2, 0.5) -- cycle; % top face
}

%\drawCube{5}{3};
\drawCube{3.85}{2.85};
\drawCube{4.35}{4.35};
\drawCube{5.85}{3.85};
\drawCube{5.35}{2.35};
%\draw[dashed,rotate=-44,blue] (1.9,5.5) ellipse (1.5 and 0.9);

    % Draw angles (transmitter)

    \draw[dashed] (0,3) -- (4.5,4.65);
    \draw[dashed] (0,3) -- (4.5,1.35);
    %\draw [<->] (1.5,2.4) arc (-5:52:1);
    %\node at (1.8, 3) {$\phi_\mathrm{TX}$};
    % Draw angles (Receiver)
    \draw[dashed] (10,3) -- (5.5,4.65);
    \draw[dashed] (10,3) -- (5.5,1.35);
    %\draw [<->] (8.9,3.4) arc (1:52:-1);
    %\node at (8.4, 3) {$\phi_\mathrm{RX}$};
    
    % Label the antennas
    \node at (0, -0.2) {transmitter};
    \node at (10,-0.2) {receiver};

    % Label the coverage area
    \node at (5,1) {cluster of scatterers};
    \node at (0, 3.8) {$N$ antennas};
    \node at (10, 3.8) {$M$ antennas};

\end{tikzpicture}}
\caption{Considered point-to-point massive MIMO system.}\label{fig:sys model}
\end{figure}

%=========================================================================
\section{Numerical Results} \label{sec:NR}
%=========================================================================

In this section, we evaluate the proposed ML data detection method in terms of SER for the considered point-to-point massive MIMO system employing 1-bit DACs at the transmitter. We assume far-field propagation and generate the channel matrix $\H$ based on the discrete physical channel model described in~\cite{Sey02}. We assume that both the transmitter and the receiver are equipped with a uniform linear array (ULA) with half-wavelength antenna spacing. The arrays are positioned such that their broadside directions are aligned. A cluster of $10^2$ scatterers is placed between the transmitter and the receiver, giving rise to as many independent propagation paths. At both ends, the scatterers are confined within an angular spread of $\frac{\pi}{6}$. The considered system is depicted in Fig.~\ref{fig:sys model}. The channels are normalized such that their elements have unit variance. The following SER results are obtained by averaging over $10^2$ independent channel and AWGN realizations, as well as $10^4$ independent data symbol vectors drawn from the 16-QAM constellation. Furthermore, for each realization of $\H$, the precoding matrix $\W$ is set to comprise the $K$ principal right eigenvectors of $\H$. Unless otherwise stated, we assume $\rho=5$~dB and $N = 128$.

\begin{figure}[t!]
\centering
\begin{subfigure}{\columnwidth}
\centering
\begin{tikzpicture}

\begin{axis}[
	width=8cm,
	height=6.5cm,
	xmin=-40, xmax=0,
	ymin=1e-4, ymax=1,
    xlabel={$\sigma^2$ [dBm]},
    ylabel={SER},
    xlabel near ticks,
    ylabel near ticks,
    xtick={-40,-30,-20,-10,0},
    xticklabels={-10,0,10,20,30},
	%log ticks with fixed point,
	ymode=log,
    %legend pos=center east,
    legend style={at={(0.99,0.01)}, anchor=south east},
	legend style={font=\scriptsize, inner sep=1pt, fill opacity=0.75, draw opacity=1, text opacity=1},
    legend cell align=left,
	grid=both,
	%minor grid style={line width=0.2pt, draw=gray!40},
	% grid style={line width=.1pt, draw=gray!20},
	%minor tick num=2,
	x label style={font=\footnotesize},
	y label style={font=\footnotesize},
	ticklabel style={font=\footnotesize},
    clip marker paths=true,
 title style={font=\scriptsize},
]

%\addplot[line width=1pt, black]
%table[x=rho_dB, y=SER1, col sep=comma] 
%{Figures/Data/SER_120deg_with_Av_EX.txt};
%\addlegendentry{\footnotesize{Case 1:}}

\addplot[very thick, black]
table [x=Delta, y=SER_ML_SVD, col sep=comma] 
{Figs/Camera_ready/SER_vs_delta_M=16_N=128_K=1.txt};
\addlegendentry{ML ($K=1$)}

\addplot[very thick, black, dashed]
table [x=Delta, y=SER_SVD, col sep=comma] 
{Figs/Camera_ready/SER_vs_delta_M=16_N=128_K=1.txt};
\addlegendentry{BLMMSE ($K=1$)}

\addplot[very thick, blue]
table [x=Delta, y=SER_ML_SVD, col sep=comma] 
{Figs/Camera_ready/SER_vs_delta_M=16_N=128_K=2.txt};
\addlegendentry{ML ($K=2$)}

\addplot[very thick, blue, dashed]
table [x=Delta, y=SER_SVD, col sep=comma] 
{Figs/Camera_ready/SER_vs_delta_M=16_N=128_K=2.txt};
\addlegendentry{BLMMSE ($K=2$)}

\addplot[very thick, red]
table [x=Delta, y=SER_ML_SVD, col sep=comma] 
{Figs/Camera_ready/SER_vs_delta_M=16_N=128_K=3.txt};
\addlegendentry{ML ($K=3$)}

\addplot[very thick, red, dashed]
table [x=Delta, y=SER_SVD, col sep=comma] 
{Figs/Camera_ready/SER_vs_delta_M=16_N=128_K=3.txt};
\addlegendentry{BLMMSE ($K=3$)}

\end{axis}

\end{tikzpicture}
\caption{$M = 16$.} \label{fig:SER4}
\end{subfigure}\par\medskip
\begin{subfigure}{\columnwidth}
\centering
\begin{tikzpicture}

\begin{axis}[
	width=8cm,
	height=6.5cm,
	xmin=-40, xmax=0,
	ymin=1e-5, ymax=1,
    xlabel={$\sigma^2$ [dBm]},
    ylabel={SER},
    xlabel near ticks,
    ylabel near ticks,
    xtick={-40,-30,-20,-10,0},
    xticklabels={-10,0,10,20,30},
	%log ticks with fixed point,
	ymode=log,
    %legend pos=center east,
    legend style={at={(0.99,0.01)}, anchor=south east},
	legend style={font=\scriptsize, inner sep=1pt, fill opacity=0.75, draw opacity=1, text opacity=1},
    legend cell align=left,
	grid=both,
	%minor grid style={line width=0.2pt, draw=gray!40},
	% grid style={line width=.1pt, draw=gray!20},
	%minor tick num=2,
	x label style={font=\footnotesize},
	y label style={font=\footnotesize},
	ticklabel style={font=\footnotesize},
    clip marker paths=true,
 title style={font=\scriptsize},
]

%\addplot[line width=1pt, black]
%table[x=rho_dB, y=SER1, col sep=comma] 
%{Figures/Data/SER_120deg_with_Av_EX.txt};
%\addlegendentry{\footnotesize{Case 1:}}

\addplot[very thick, black]
table [x=Delta, y=SER_ML_SVD, col sep=comma] 
{Figs/Camera_ready/SER_vs_delta_M=64_N=128_K=1.txt};
\addlegendentry{ML ($K=1$)}

\addplot[very thick, black, dashed]
table [x=Delta, y=SER_SVD, col sep=comma] 
{Figs/Camera_ready/SER_vs_delta_M=64_N=128_K=1.txt};
\addlegendentry{BLMMSE ($K=1$)}

\addplot[very thick, blue]
table [x=Delta, y=SER_ML_SVD, col sep=comma] 
{Figs/Camera_ready/SER_vs_delta_M=64_N=128_K=2.txt};
\addlegendentry{ML ($K=2$)}

\addplot[very thick, blue, dashed]
table [x=Delta, y=SER_SVD, col sep=comma] 
{Figs/Camera_ready/SER_vs_delta_M=64_N=128_K=2.txt};
\addlegendentry{BLMMSE ($K=2$)}

\addplot[very thick, red]
table [x=Delta, y=SER_ML_SVD, col sep=comma] 
{Figs/Camera_ready/SER_vs_delta_M=64_N=128_K=3.txt};
\addlegendentry{ML ($K=3$)}

\addplot[very thick, red, dashed]
table [x=Delta, y=SER_SVD, col sep=comma] 
{Figs/Camera_ready/SER_vs_delta_M=64_N=128_K=3.txt};
\addlegendentry{BLMMSE ($K=3$)}
\end{axis}

\end{tikzpicture}
\caption{$M = 64$.} \label{fig:SER5}
\end{subfigure}\par\medskip
\caption{SER versus dithering power, with $\rho = 5$~dB, $N = 128$.} \label{fig:SER4-5}
\end{figure}
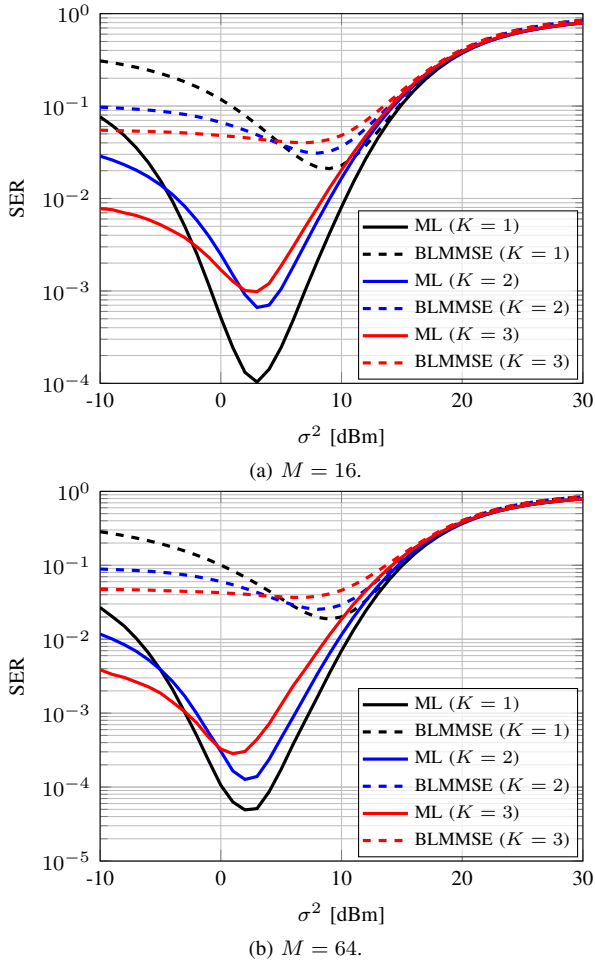

Fig.~\ref{fig:SER4-5} compares the proposed ML method with BLMMSE with $M \in \{ 16, 64 \}$ and $K \in \{ 1, 2, 3\}$. We observe that all the SER curves feature an optimal operating point in terms of dithering power. On the one hand, for low $\sigma^2$, the data symbols with the same phase give rise to approximately the same quantized precoded signal: this is because the precoded signal becomes highly distorted by the 1-bit DACs. On the other hand, for high $\sigma^2$, the Gaussian dithering is dominant. Fig.~\ref{fig:SER4} shows that, with $M = 16$, the proposed ML method can provide gains of more than two orders of magnitude in terms of SER compared with BLMMSE for $K=1$. This is because BLMMSE does not consider the statistics of the received signal and only attempts to minimize the MSE for Gaussian data symbols. Both the proposed ML method and BLMMSE improve as $K$ increases when the dithering power is low. Indeed, in the absence of dithering, the increased interference from additional data streams produces a beneficial scrambling of the precoded signal across the transmit antennas before the 1-bit quantization. In contrast, when the dithering power is optimized, stronger inter-stream interference leads to higher SER. Fig.~\ref{fig:SER5} extend the insights of Fig.~\ref{fig:SER4} to the case of $M=64$. Comparing Fig.~\ref{fig:SER4} and Fig.~\ref{fig:SER5}, we observe that the proposed ML method attains approximately a $6 \times$ SER improvement for $K = 2$ when increasing the number of antennas from $M = 16$ to $M = 64$, while BLMMSE shows almost no gain under the same conditions.

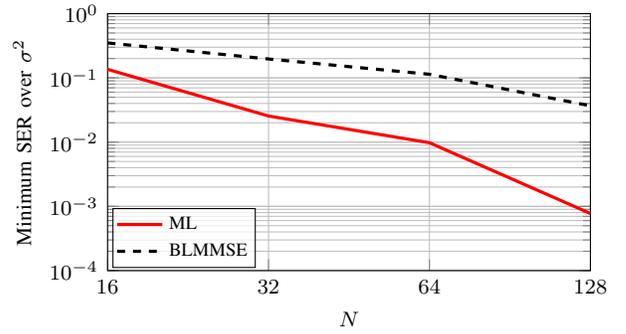
\begin{figure}[t!]
\centering
\begin{tikzpicture}

\begin{axis}[
	width=8cm,
	height=5cm,
	xmin=16, xmax=128,
	ymin=1e-4, ymax=1,
    xlabel={$N$},
    ylabel={Minimum SER over $\sigma^2$},
    xlabel near ticks,
	ylabel near ticks,
    xtick={16,32,64,128},
    ytick={0.0001,0.001,0.01,0.1,1},
    yticklabels={$10^{-4}$,$10^{-3}$,$10^{-2}$,$10^{-1}$,$10^{0}$},
	log ticks with fixed point,
	ymode=log,
   legend pos=north east,
    legend style={at={(0.01,0.01)}, anchor=south west},
	legend style={font=\scriptsize, inner sep=1pt, fill opacity=0.75, draw opacity=1, text opacity=1},
    legend cell align=left,
	grid=both,
%	major grid style={line width=0.5pt, draw=gray!40},
%	grid style={line width=.2pt, draw=gray!20},
%	minor tick num=4,
	x label style={font=\footnotesize},
	y label style={font=\footnotesize},
	ticklabel style={font=\footnotesize},
    clip marker paths=true,
    title style={font=\scriptsize},
    xmode=log,
	log basis x={2},
]

\addplot[very thick, red]
table [x=N, y=SERmin_ML, col sep=comma] 
{Figs/Camera_ready/minSER,ML,BLMMSE,N.txt};
\addlegendentry{ML}

\addplot[very thick, dashed, black]
table [x=N, y=SERmin_BLMMSE, col sep=comma] 
{Figs/Camera_ready/minSER,ML,BLMMSE,N.txt};
\addlegendentry{BLMMSE}

\end{axis}

\end{tikzpicture}
\caption{Minimum  SER versus number of transmit antennas, with $\rho = 5$~dB, $K = 3$, and $M = 16$.}\label{fig:min_SER}
\end{figure}

Considering $K = 3$ and $M = 16$, Fig.~\ref{fig:min_SER} demonstrates the impact of $N$ on the minimum SER over $\sigma^2$. As $N$ increases, the SER obtained by both the proposed ML method and BLMMSE decreases; however, the decrease is steeper for the proposed ML method. Considering $\sigma^2 = 2$~dBm, $K = 3$, and $M = 16$, Fig.~\ref{fig:SER6} illustrates the effect of the SNR on the SER obtained by the proposed ML method and BLMMSE. As the SNR increases, the SER of the proposed ML method improves significantly, whereas that of BLMMSE remains approximately constant. However, both the proposed ML method and BLMMSE exhibit performance saturation at high SNRs, since increasing the SNR cannot resolve the quantization distortion at the transmitter.

\begin{figure}[t!]
\centering
\begin{tikzpicture}

\begin{axis}[
	width=8cm,
	height=5cm,
	xmin=0, xmax=40,
	ymin=1e-5, ymax=1e-1,
    xlabel={$\rho$~[dB]},
    ylabel={SER},
    xlabel near ticks,
	ylabel near ticks,
    xtick={0,10,20,30,40},
    ytick={0.00001,0.0001,0.001,0.01,0.1,1},
    yticklabels={$10^{-5}$,$10^{-4}$,$10^{-3}$,$10^{-2}$,$10^{-1}$,$10^{0}$},
	log ticks with fixed point,
	ymode=log,
   legend pos=north east,
    legend style={at={(0.01,0.01)}, anchor=south west},
	legend style={font=\scriptsize, inner sep=1pt, fill opacity=0.75, draw opacity=1, text opacity=1},
    legend cell align=left,
	grid=both,
%	major grid style={line width=0.5pt, draw=gray!40},
%	grid style={line width=.2pt, draw=gray!20},
%	minor tick num=4,
	x label style={font=\footnotesize},
	y label style={font=\footnotesize},
	ticklabel style={font=\footnotesize},
    clip marker paths=true,
    title style={font=\scriptsize},
]

\addplot[very thick, red]
table [x=rho_dB, y=SER_ML_SVD, col sep=comma] 
{Figs/Camera_ready/SER_vs_rho_M=16_N=128_K=3.txt};
\addlegendentry{ML}

\addplot[very thick,dashed, black]
table [x=rho_dB, y=SER_SVD, col sep=comma] 
{Figs/Camera_ready/SER_vs_rho_M=16_N=128_K=3.txt};
\addlegendentry{BLMMSE}
\end{axis}

\end{tikzpicture}
\caption{SER versus SNR, with $\sigma^2 = 2$~dBm, $N = 128$, $K = 3$, and $M = 16$.}\label{fig:SER6}
\end{figure}
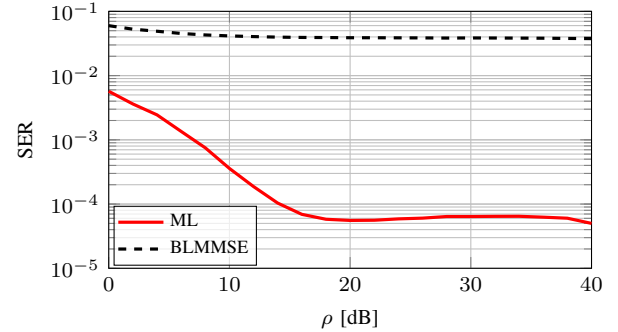

\begin{figure*}[t!]
\setcounter{equation}{25}
\begin{align}
\C_{\tilde{\n}'}^{(\mathrm{A},\mathrm{B})}\!= \!\rho\mathbb{E}\bigg[\mathrm{A}[\H\p_{\textrm{d}}]\mathrm{B}[\H\p_{\textrm{d}}]^\tran\!+ \!\mathrm{A}[\H\G(\x)\d]\mathrm{B}[\H\p_{\textrm{d}}]^\tran\!+ \!\mathrm{A}[\H\p_{\textrm{d}}]\mathrm{B}[\H\G(\x)\d]^\tran\bigg]\!+ \!\frac{\rho\sigma^2}{2}\Lambda_{\mathrm{A},\mathrm{B}}\Big(\H\G(\x)\G(\x)^\herm \H^\herm\Big)
\! +\!\frac{\delta_{\mathrm{A},\mathrm{B}}}{2}\I_M, \label{eq:Cnt_Re}
\end{align}
\vspace{-2mm}
\begin{align}
\label{eq:Cqx_re}
\mathbb{E}_{\p_{\textrm{d}}}\big[\mathrm{A}[\p_{\textrm{d}}]\mathrm{B}[\p_{\textrm{d}}]^\tran\big] &= \C_{\x_{\textrm{q}}}^{(\mathrm{A},\mathrm{B})} - {\P_1}^{(\mathrm{A},\mathrm{B})}  - {\P_2}^{(\mathrm{A},\mathrm{B})} + \frac{\sigma^2}{2}\Lambda_{\mathrm{A},\mathrm{B}}\Big(\G(\x)\G(\x)^\herm\Big) + \mathrm{A}[\G(\x)\x]\mathrm{B}[\G(\x)\x]^\tran \in \mathbb{R}^{N\times N}
\end{align}
\begin{align} \label{eq:Exq_n Exq_m}
[\C_{\x_{\textrm{q}}}^{( \mathrm{A}, \mathrm{B})}]_{n,m} \triangleq\mathbb{E}_{\x_{\textrm{q}}}\big[\mathrm{A}[\x_{\textrm{q}}]\mathrm{B}[\x_{\textrm{q}}]^\tran \big]_{n,m} & =
    \begin{cases}
    \frac{\eta}{2}\big(1 - \Phi \big(\frac{\mathrm{A}[x_n]}{\sigma} \big)\Phi \big(\frac{\mathrm{B}[x_m]}{\sigma} \big)\big)\delta_{\mathrm{A},\mathrm{B}} + \frac{\eta}{2}\Phi \big(\frac{\mathrm{A}[x_n]}{\sigma} \big)\Phi \big(\frac{\mathrm{B}[x_m]}{\sigma} \big)& \ \textrm{if}~m=n, \\
    \frac{\eta}{2} \Phi \big(\frac{\mathrm{A}[x_n]}{\sigma} \big)\Phi \big(\frac{\mathrm{B}[x_m]}{\sigma} \big) & \ \textrm{otherwise}.
    \end{cases} 
\end{align}
\hrulefill
\end{figure*}

%=========================================================================
\section{Conclusions} \label{sec:Contr_Full}
%=========================================================================

In this paper, we proposed a new ML data detection method for a point-to-point massive MIMO system with 1-bit DACs at the transmitter, where the precoded signal is supplemented with dithering before the 1-bit quantization. In this regard, we derived the mean and covariance matrix of the received signal, where symbol-dependent LMMSE estimation is utilized to efficiently linearize the transmitted signal. Lastly, we evaluated the impact of the dithering power, the number of receive antennas and data streams, and the SNR on the system's performance. Numerical results showed that the proposed ML method can provide gains of more than two orders of magnitude in terms of SER compared with conventional data detection based on soft estimation. Future work will consider imperfect CSI and explore a low-complexity variant of the proposed ML data detection method based on near ML approaches.

%Things to mention about future work:
%\begin{itemize}
%\item Include channel estimation and imperfect CSI
%\item Reduce complexity (search space)
%\item Optimization of the dithering level
%\item Optimization of the precoding matrix
%\item Extension to doubly 1-bit quantized systems
%\end{itemize}

%=========================================================================
\section*{Appendix}
%=========================================================================

Here, we derive the mean and covariance of $\tilde{\n}'$ defined in Section~\ref{sec:LQP&DD_D}. Throughout the Appendix, $\mathrm{A}[\cdot]$ and $\mathrm{B}[\cdot]$ can be either $\Re[\cdot]$ or $\Im[\cdot]$. Moreover, unless otherwise stated, all the expectations are taken over $\d$, $\p_{\textrm{d}}$, and $\z$, conditioned on a given $\x$. The mean of $\tilde{\n}'$ is given by
\setcounter{equation}{19}
\begin{align}
\boldsymbol{\mu}_{\tilde{\n}'}& = \big[\mathbb{E}\big[\Re[\tilde{\n}]\big], \mathbb{E}\big[\Im[\tilde{\n}]\big]\big]^\tran,\label{eq:mu_n'} 
\end{align}
with
\begin{align}
\mathbb{E}\big[\Re[\tilde{\n}]\big] &= \sqrt{\rho} \big(\Re[\H]\mathbb E\big[\Re[\p_{\textrm{d}}]\big] - \Im[\H]\mathbb E\big[\Im[\p_{\textrm{d}}]\big] \big), \label{eq:E[Re[zt]]}
     \\ \mathbb {E}\big[\Im[\tilde{\n}]\big] &= \sqrt{\rho} \big(\Re[\H]\mathbb E\big[\Im[\p_{\textrm{d}}]\big] + \Im[\H]\mathbb E\big[\Re[\p_{\textrm{d}}]\big] \big)\label{eq:E[Im[zt]]}
\end{align}
and
\begin{align}
    \mathbb E\big[\mathrm{A}[\p_{\textrm{d}}]\big] &= \mathbb E\big[\mathrm{A}[\x_{\textrm{q}}]\big] -  \mathrm{A} \big[\G(\x)\x \big] \\& =
    \sqrt{\frac{\eta}{2}} \Phi\bigg(\frac{\mathrm{A}[\x]}{\sigma}\bigg) - \mathrm{A} \big[\G(\x)\x \big]. \label{eq:Eqx_A}
\end{align} 
The second-order moment of $\tilde{\n}'$, denoted as $\C_{\tilde{\n}'}\triangleq \mathbb{E} \big[\tilde{\n}'{(\tilde{\n}')}^\tran \big]$, is given by 
\begin{align}\label{eq:Cn't}
    \C_{\tilde{\n}'}&=\begin{bmatrix}
        \C_{\tilde{\n}'}^{(\Re,\Re)} & \C_{\tilde{\n}'}^{(\Re,\Im)}\\
        (\C_{\tilde{\n}'}^{(\Re,\Im)})^\tran & \C_{\tilde{\n}'}^{(\Im,\Im)}
    \end{bmatrix} \in \mathbb{R}^{2M\times 2M},
    \end{align}
with specific term in \eqref{eq:Cnt_Re}--\eqref{eq:Exq_n Exq_m} at the top of the page. In \eqref{eq:Cnt_Re}, the cross-correlation matrix between $\d$ and $\p_{\textrm{d}}$ is given by
\setcounter{equation}{28}
\begin{align}
    \nonumber\mathbb{E}_{\d,\p_{\textrm{d}}}\big[ \mathrm{A}[\d]\mathrm{B}[\p_{\textrm{d}}]^\tran\big] &= \C_{\x_{\textrm{d}}\x_{\textrm{q}}}^{( \mathrm{A}, \mathrm{B})} - \mathbb{E}\big[\mathrm{A}[\d]\mathrm{B} \big[\G(\x)\d \big]^\tran\big] \\
    & \phantom{=} \ - \mathrm{A}[\x]\mathbb{E}\big[\mathrm{B}[\x_{\textrm{q}}]\big]^\tran, \label{eq:Edqx_AB}
\end{align}
with $\mathbb{E}\big[\mathrm{B}[\x_{\textrm{q}}]\big]$ given in \eqref{eq:Eqx_A}, whereas $\mathbb{E}\big[\mathrm{A}[\d]\mathrm{B} \big[\G(\x)\d \big]^\tran\big]$ is straightforward to obtain since $\mathbb{E}\big[\mathrm{A}[\d]\mathrm{B}[\d]^\tran\big] = \delta_{\mathrm{A},\mathrm{B}}\, \sigma^2\I_N$. Moreover, in \eqref{eq:Cqx_re}, we defined
\begin{align} \label{eq:P1}
    {\P_1}^{(\mathrm{A},\mathrm{B})} &\triangleq \mathbb{E}_{\d,\x_{\textrm{q}}}\big[ \mathrm{A}[\G(\x)\x_{\textrm{d}}]\mathrm{B}[\x_{\textrm{q}}]^\tran \big], \\\label{eq:P2}
    {\P_2}^{(\mathrm{A},\mathrm{B})} &\triangleq \mathbb{E}_{\d,\x_{\textrm{q}}}\big[\mathrm{A}[\x_{\textrm{q}}] \mathrm{B}[\G(\x)\x_{\textrm{d}}]^\tran \big],
\end{align}
where the $(n,m)$th element of $\mathbb{E}_{\d,\x_{\textrm{q}}}\big[ \mathrm{A}[\x_{\textrm{d}}]\mathrm{B}[\x_{\textrm{q}}]^\tran \big]$ is given in \eqref{eq:Exd_n Exq_m (App)}. Finally, the covariance matrix of $\tilde{\n}'$ is obtained as
\begin{align}
    \mathbf{\Sigma}_{\tilde{\n}'} = \C_{\tilde{\n}'} - \boldsymbol{\mu}_{\tilde{\n}'}\boldsymbol{\mu}_{\tilde{\n}'}^\tran. \label{eq:Sigma_n'}
\end{align}

\addcontentsline{toc}{chapter}{References}
\bibliographystyle{IEEEtran}
\bibliography{refs_abbr,refs}

\end{document}